\documentclass[twocolumn,showpacs,preprintnumbers,amsmath,amssymb]{revtex4}
\usepackage{graphicx}% Include figure files
\usepackage{dcolumn}% Align table columns on decimal point
\usepackage{bm}% bold math

\begin{document}

%\preprint{APS/123-QED}

%\title{Derivation of the Bohr ground-state atomic radius from spin-orbit interaction and hypothesis of angular momentum %constancy}% Force line breaks with \\

%\title{Spin-orbit interaction dynamics of atomic hydrogen in classical electrodynamics}% Force line breaks %with \\

\title{On the Bohr radius relationship to spin-orbit interaction, spin magnitude, and Thomas precession}% Force line breaks with \\

\author{David C. Lush } 
\affiliation{%
 d.lush@comcast.net \\}%


\date{\today}% It is always \today, today,

\begin{abstract} 

The dynamics of the spin-orbit interaction in atomic hydrogen are studied in a classical electrodynamics-like setting.  A Rutherfordian atomic model is used assuming a circular electron orbit, without the quantum principle as imposed arbitrarily in the Bohr model, but with an \textit{ad hoc} incorporation in the electron of intrinsic spin and associated magnetic dipole moment.   Analyzing the motions of the electron spin and orbital angular momenta, it is found that in the presence of Thomas precession, the total angular momentum averaged over the orbit is not generally a constant of the motion.  It is noted this differs from the finding of Thomas in a similar assessment of 1927, and the reason for this difference is provided.  It is found that although the total orbit-averaged angular momentum is not a constant of the motion, it precesses around a fixed axis similarly to the precession of the total angular momentum vector seen in spin-orbit coupling in quantum theory.  The magnitude of the angular velocity of the total orbit-averaged angular momentum is seen to vanish only when the spin and orbital angular momenta are antiparallel and their mutual precession frequencies equate. It is then found, there is a unique radius where the mutual precession frequencies equate.  Assuming the electron magnetic moment is the Bohr magneton, and an electron g-factor of two, this radius corresponds to where the orbital angular momentum is the reduced Planck's constant. The orbit radius for stationary total angular momentum for the circular orbit model with nonzero orbital angular momentum is thus the ground-state radius of the Bohr model.     
   
\end{abstract} 

% \begin{keywords}
%   Bohr Correspondence Principle, Spin-orbit Interaction, intrinsic angular momentum,
%   Larmor precession, Thomas precession, Thomas factor,
%   orbital angular momentum, quantum theory
% \end{keywords}

%\end{titlepage}

%\pagebreak
\pacs{41.20.-q, 31.15Gy, 45.05.+x}% PACS, the Physics and Astronomy
%\twocolumn
%\addcontentsline{hydrogen}{section}{Introduction}
% 45+

\maketitle

\section{Introduction}

Bohr's model of atomic hydrogen \cite{Bohr:1914} is a modification of the prior Rutherford model with an \textit{ad hoc} introduction of a quantum principle.  The quantum principle in the Bohr model is that stable electron orbits are those with angular momentum in integer multiples of the reduced Planck's constant, \( \hbar \). This model was singularly successful in explaining the low-resolution emission spectrum of hydrogen and, when extended by Sommerfeld from circular to elliptical orbits, even the fine structure.  The general approach proved difficult to extend to multi-electron atoms, however, and was soon superceded by the more generally applicable theories of Heisenberg and Shr\"odinger. Apart from their general unsuitability for use for either many-electron atoms or molecules, both the Bohr and Sommerfeld models suffer a more fundamental problem due to disagreement with modern understanding of the angular momentum of the hydrogenic ground states.  The Bohr and Sommerfeld models will have orbital angular momentum of \(\hbar\) in these states, whereas, it is now understood that the quantum mechanical ground states, as well as all other ``s'' states at elevated energies, have total angular momentum eigenvalues consistent only with vanishing orbital angular momentum.  

It is important to the current work therefore that recently the Sommerfeld model has been reinterpreted by Bucher \cite{BucherY8} so that it provides much better agreement with modern quantum theory of hydrogenic atoms and the hydrogen molecule ion. By ruling out the circular orbit while admitting the possibility of a zero orbital angular momentum state termed by Bucher the Coulomb oscillator \cite{BucherY6}, close correspondence and agreement with modern quantum theory is obtained within the scope of applicability of the Sommerfeld model. This agreement extends even to the hyperfine splitting, from which it must be recognized that the electron in the ground state of atomic hydrogen must have some overlap of position with the nucleus.  

In the present work we begin as did Bohr and Sommerfeld with the Rutherford atomic model consisting of point-like charged particles moving in closed Galilean orbits under Coulomb attraction.  It is then investigated, within the confines of this model, what are the dynamical effects of the fact that the electron possesses in addition to mass and charge a property of intrinsic spin.  

The existence of intrinsic spin and its associated magnetic moment, unknown at publication of Bohr's model, may naturally be considered of potential applicability due to the spin magnitude directly involving Planck's constant. That quantum behavior is a direct consequence of intrinsic spin is proposed by Hestenes \cite{Hestenes:1990}, who observes that the dependence of both the uncertainty relations and the intrinsic spin on Planck's constant ``cries out for some explanation'' \cite{Hestenes:1979}. Hestenes proposes that the spin magnitude itself participating in the dynamics may provide the explanation for nonradiative states via resonance phenomena \cite{Hestenes:1991}.  Recent work by De Luca \cite{deLuca:2006} on the relativistic Lorentz-Dirac electrodynamics of two point particles supports Hestenes' contention.  Exactly solving even the two-body problem is an ongoing effort requiring development of new mathematical techniques \cite{deLuca:2007} that take full account of delay and the self force, however.  A simplified approach is thus desirable. 

Although the present work, neglecting as it does both delay and radiation reaction, and with the spin magnitude constant and unable to participate in the dynamics, must be regarded as at best an effective theory or ``toy'' model, we argue it is worth examination and extension nonetheless, due to its initial successes as demonstrated herein.  These include a direct linking of the Bohr radius value, and thus the atomic scale, to the electron intrinsic spin and magnetic moment magnitudes, a new interpretation of the principal quantum number as the ratio of spin to orbit precession frequencies, and a quasiclassical-physics explanation for the Bohr Correspondence Principle.  We also propose a direct linkage between the necessity of quantization and Thomas precession.

Finally, it is to be noted that Llewellyn Thomas, in his landmark paper of 1927 \cite{Thomas1927}, considered an atomic model similar to that used here. In addition to deriving the ``relativity precession'' that now bears his name, he considered the similar model to that used here as part of his application of the Thomas precession to explaining the anomalous factor of one-half in the Zeeman effect.  Therefore it is relevant to understanding the present work to recognize that Thomas obtained essentially an opposite result to that of the present analysis, concerning the conservation of the secular total angular momentum consisting of the orbit-averaged electron orbital and intrinsic angular momenta.  However, it can be seen \cite{LushY9b}, Thomas's analysis did not take into account the ``hidden momentum'' of a magnetic dipole in an electric field.  This is understandable since the need for such accounting was not recognized until the latter half of the twentieth century \cite{Furry1969},\cite{Hnizdo1992},\cite{jcksn:classelec3}.  Due to this omission Thomas concluded, correctly for his model and the electrodynamics of his time, but ultimately incorrectly, that the total secular mechanical angular momentum is a constant of the motion generally for circular as well as elliptical orbits.  The conclusion of the present analysis however is that for nonzero orbital angular momentum the secular total angular momentum is not a constant of the motion except for antiparallel spin and orbital angular momenta, and further assuming a circular orbit, for an orbital radius equal to the ground-state radius of the Bohr model.  For the circular orbit this corresponds to an orbital angular momentum of \(\hbar\).  The restriction to circular orbits of the present analysis furthermore is for convenience only and not thought to be essential.  An analysis to be presented later \cite{LushY9c} has confirmed that allowing for elliptical orbits, equation of the spin and secular orbital angular momentum precession frequencies provides a quasiclassical basis for Sommerfeld's azimuthal quantum rule.

\section{Angular Momentum Equations of Motion in the Laboratory Frame}

In this section the equations of motion of the spin and orbital angular momenta are determined for the laboratory frame.  The forces causing the  precessional motions will be calculated in the electron rest frame, in which the proton orbits around the electron, however.  The torques on the electron spin and proton orbit around the electron in the electron rest frame will then be evaluated to determine the motions of the angular momentum vectors in the laboratory frame.

\subsection{Definition and Use of the Electron Body Frame}

The analysis will consider what are the electrodynamic forces present in a coordinate frame comoving with the electron, called the electron rest or body frame.  While this is a common usage as a convenience in textbook calculations of spin-orbit interaction effects and energies, we will use it in a somewhat different and arguably expanded sense in the present work.  All electron body frame forces will be transformed back to the laboratory frame prior to consideration of their dynamical effect.  Dynamical equations will be written in the laboratory frame exclusively.       

A coordinate frame comoving with the electron as it orbits around the proton under Coulomb attraction undergoes a rotation with angular velocity \cite{jcksn:classelec}

\begin{equation}
\mbox{\boldmath$\omega$}_{\text{T}} =
\frac{\gamma^2}{\gamma+1}\frac{\mbox{\boldmath$a$} \times
\mbox{\boldmath$v$} } {c^2} = - \omega_{\text{T}} \hat{\mbox{\boldmath$L$}}
\label{Thomas_freq1}
\end{equation}

where \(\mbox{\boldmath$v$} \) and \( \mbox{\boldmath$a$} \) are the electron velocity and acceleration, \( c\) is the speed of light, and \(\gamma = \sqrt{1/(1-(v/c)^2)} \).  The existence of this rotation, now known as Thomas precession, and its relevance to the spin-orbit interaction was first recognized by Thomas \cite{Thomas1927}, and used by him to explain the factor of \(1/2\) in the anomalous Zeeman effect.  

The formula for \( \omega_{\text{T}} \) for circular orbits as will be required is worked out in the Appendix.

\subsection{Transformation from the Electron Body Frame to the Laboratory Frame}

The orbital angular momentum in the laboratory and electron rest frames may be represented \cite{gldstn:classmech_oc133} as

\begin{equation}
\left( \frac{d \mbox{\boldmath$L$}}{dt} \right)_{\text{lab}} = \left(
\frac{d \mbox{\boldmath$L$}}{dt} \right)_{\text{elec}} +
\mbox{\boldmath$\omega$}_{\text{T}} \times \mbox{\boldmath$L$}
\label{Ltransform}
\end{equation}

The second term on the right vanishes due to the parallelness of \( \mbox{\boldmath$\omega$}_{\text{T}} \) and \( \mbox{\boldmath$L$} \). Now, \( \mbox{\boldmath$L$} \) in the electron rest frame is due to the orbital motion of the proton around the electron and neglecting the (small) effect of the Thomas precession is the proton-to-electron mass ratio times the electron orbital angular momentum in the laboratory frame.  Also, it will be straightforward to calculate the torque on the proton orbit in the electron rest frame due to the proton velocity through the intrinsic magnetic field of the electron. However, due to the acceleration of the electron frame we must not take for granted that the time derivative of the orbital angular momentum is equal to the torque in that frame.  To assess this, let us evaluate

\begin{eqnarray}
\left(\frac{d \mbox{\boldmath$L$}}{dt} \right)_{\text{elec}} && =
\left(\frac{d (\mbox{\boldmath$r$}_p \times m_p \mbox{\boldmath$v$}_p )}{dt} \right)_{\text{elec}} \nonumber\\
&& =
m_p \left( \mbox{\boldmath$v$}_p \times \mbox{\boldmath$v$}_p + \mbox{\boldmath$r$}_p \times  \mbox{\boldmath$a$}_p  \right)_{\text{elec}} \nonumber\\
 && =
\left(\mbox{\boldmath$r$}_p \times  \mbox{\boldmath$F$}_p  \right)_{\text{elec}} 
\equiv \mbox{\boldmath$\tau$}_L
\label{Lelec3}
\end{eqnarray}

where the subscripts \(p\) indicate proton quantities in the electron rest frame and \( \mbox{\boldmath$\tau$}_L \) may be identified as the torque on the proton orbit in the electron rest frame.  Therefore, the time rate of change of the orbital angular momentum in the laboratory frame is identically equal to the torque experienced by the orbiting proton in the electron rest frame.

Similarly, for the electron spin,

\begin{equation}
\left( \frac{d \mbox{\boldmath$s$}}{dt} \right)_{\text{lab}} = \left(
\frac{d \mbox{\boldmath$s$}}{dt} \right)_{\text{elec}} +
\mbox{\boldmath$\omega$}_{\text{T}} \times \mbox{\boldmath$s$}
\label{spintransform}
\end{equation}

However, the spin will not in general be aligned with the orbital angular momentum and Thomas angular velocity, so the form of (\ref{spintransform}) needed is

\begin{equation}
\left( \frac{d \mbox{\boldmath$s$}}{dt} \right)_{\text{lab}} = \mbox{\boldmath $\tau$}_s +
\mbox{\boldmath$\omega$}_{\text{T}} \times \mbox{\boldmath$s$}
\label{spintransform2}
\end{equation}

where \( \mbox{\boldmath $\tau$}_s \) is the torque on the spin evaluated in the electron rest frame.   We are now equipped to evaluate the orbital angular momentum and electron spin precessional angular velocities in the laboratory frame.

\section{Motions of the Spin and Orbit Angular Momenta}

The derivation of this section proceeds simply as follows.  The motion of the orbit and spin angular momenta are calculated in the laboratory frame for the circular orbit and as a function of electron-proton separation.    

\subsection{Motion of the Orbital Angular Momentum}

In the electron rest frame, the magnetic field due to the electron intrinsic magnetic moment causes a force on the orbiting proton according to the proton velocity and electrical charge.  The magnetic field at a point outside the source region is given in terms of the magnetic moment \(\mbox{\boldmath$m$}\) of the source as  \cite{jcksn:classelec_B_mm}  

\begin{equation}
\mbox{\boldmath$B$} = \frac{3\mbox{\boldmath$n$} \left(
\mbox{\boldmath$n$} \cdot \mbox{\boldmath$m$}   \right)  -
\mbox{\boldmath$m$}}{R^3}
\label{Bduetoelecmm2}
\end{equation}

where \( \mbox{\boldmath$n$}=\mbox{\boldmath$r$}/R \) is a unit vector in the direction from the source to the field point, \( \mbox{\boldmath$r$}\) here and throughout is a vector from the proton to the electron, and \( R \equiv |\mbox{\boldmath$r$}| \).

The electron intrinsic magnetic moment will be  represented as

\begin{equation}
\mbox{\boldmath$\mu$} = -\frac{ge}{2 m_e c} \mbox{\boldmath$s$}=
-\frac{ges}{2 m_e c}
\hat{\mbox{\boldmath$s$}}
\label{defofintrnsicmm}
\end{equation}

where \( s \) is the spin angular momentum magnitude  and
\(\hat{\mbox{\boldmath$s$}}\) is a unit-magnitude orientation
vector, and where \(e\) is the electron charge magnitude, \( m_e\) is the  electron rest mass, \(c\) is the speed of light, and where \(g\) the electron gyromagnetic factor.  

It is commonly accepted that \(s = \hbar/2 \) where \(\hbar = h/(2\pi)\) is the reduced Planck's constant, and that \(g \approx 2\). 

The torque on the proton orbit around the electron is then

\begin{equation}
\mbox{\boldmath$\tau$}_L = \mbox{\boldmath$r$}_p \times
\mbox{\boldmath$F$}_p = \mbox{\boldmath$r$}_p \times \left(
\frac{e}{c}\mbox{\boldmath$v$}_p \times \frac{3\mbox{\boldmath$n$}
\left( \mbox{\boldmath$n$} \cdot \mbox{\boldmath$\mu$}   \right) -
\mbox{\boldmath$\mu$}}{R^3} \right)
\label{torqueporbdtemm}
\end{equation}

with \( \mbox{\boldmath$n$} = \mbox{\boldmath$r$}_p / R \) here, and \( \mbox{\boldmath$v$}_p \) is the proton velocity as measured in the electron rest frame.  

The vector triple product of (\ref{torqueporbdtemm}) involving \(\mbox{\boldmath$\mu$}\) can be expanded as

\begin{equation}
\mbox{\boldmath$r$}_p \times (\mbox{\boldmath$v$}_p \times \mbox{\boldmath$\mu$}) = (\mbox{\boldmath$r$}_p \cdot\mbox{\boldmath$\mu$})\mbox{\boldmath$v$}_p - (\mbox{\boldmath$r$}_p \cdot \mbox{\boldmath$v$}_p)\mbox{\boldmath$\mu$} 
= R(\mbox{\boldmath$n$} \cdot\mbox{\boldmath$\mu$})\mbox{\boldmath$v$}_p \label{vtpident2}
\end{equation}

since \(  \mbox{\boldmath$n$} = \mbox{\boldmath$r$}_p / R \) here, and recognizing that the proton velocity and position vectors are orthogonal for the circular orbit so that the second term in the center vanishes.  Similarly the vector triple product of Eq.(\ref{torqueporbdtemm}) involving \(  \mbox{\boldmath$n$} \)  yields that \( \mbox{\boldmath$r$}_p \times (\mbox{\boldmath$v$}_p \times \mbox{\boldmath$n$}) = R \mbox{\boldmath$v$}_p \) and so Eq. (\ref{torqueporbdtemm}) becomes

\begin{equation}
\mbox{\boldmath$\tau$}_L = \frac{2e}{cR^2}
(\mbox{\boldmath$n$} \cdot \mbox{\boldmath$\mu$}) \mbox{\boldmath$v$}_p 
\label{torqueporbdtemm3}
\end{equation}

So, for any spin orientation other than parallel to the orbital angular momentum vector, the torque is time-varying during the orbit.  It will therefore be of interest to compute the average torque over the course of an orbit.  The spin and orbital angular momentum vectors precess very slowly compared to an orbital period so it is reasonable to treat their relative orientation as fixed during an orbit.  It will be convenient to choose for the electron rest frame Cartesian coordinate axes with directions \( \hat{\mbox{\boldmath$x$}}, \hat{\mbox{\boldmath$y$}},\hat{\mbox{\boldmath$z$}} \) with origin at the electron and where the orbital angular momentum \( \hat{\mbox{\boldmath$L$}} \) is in the \( \hat{\mbox{\boldmath$z$}} \) direction.  We suppose that in general the electron spin is not aligned with \( \hat{\mbox{\boldmath$L$}} \) and choose the \( \hat{\mbox{\boldmath$x$}}\) direction to be aligned with the projection of \( \mbox{\boldmath$s$} \) into the orbital plane.  We also choose the time origin so that the proton at \(t=0\) is in the \( \hat{\mbox{\boldmath$x$}}\) direction. Then over a time scale where the precessional motion of the spin may be neglected \( (\mbox{\boldmath$n$} \cdot \mbox{\boldmath$\mu$}) =  \mu_\perp \cos(\omega t) \), and (\ref{torqueporbdtemm3}) can be rewritten as

\begin{equation}
\mbox{\boldmath$\tau$}_L = \frac{2 e }{cR^2}\mu_\perp \cos(\omega t)  \mbox{\boldmath$v$}_p 
\label{torqueporbdtemm5}
\end{equation}

where \(\mu_{\perp}\) is the electron intrinsic magnetic moment component in the orbital plane and \( \omega \) here strictly is the orbital frequency of the proton around the electron in the electron rest frame.  Expanding the velocity we have

\begin{equation}
\mbox{\boldmath$\tau$}_L = \frac{2 e \mu_\perp \cos(\omega t)}{cR^2}  v_p (-\sin(\omega t)\hat{\mbox{\boldmath$x$}}  +   \cos(\omega t)\hat{\mbox{\boldmath$y$}})
\label{torqueporbdtemm5a}
\end{equation}

Integrating over an orbital period \( T = 2\pi/\omega\) and dividing by \(T\) to obtain the average, the \(x\) component with \( \sin(\omega t) \cos(\omega t) \) vanishes and the \( \cos^2(\omega t) \) factor on the \(y\) component contributes a factor of a half and so 

\begin{equation}
\langle \mbox{\boldmath$\tau$}_L \rangle = \frac{e\mu_\perp}{cR^2}  \frac{e}{\sqrt{m_e R}}\hat{\mbox{\boldmath$y$}}
\label{torqueporbdtemm6}
\end{equation}

where angle brackets indicate the average over a turn of the orbit.  But

\begin{equation}
\mu_\perp\hat{\mbox{\boldmath$y$}} = -\mu_\perp (\hat{\mbox{\boldmath$x$}} \times \hat{\mbox{\boldmath$L$}}) \equiv -\mbox{\boldmath$\mu$}_{\perp}\times \hat{\mbox{\boldmath$L$}}
\label{muperpcrossL}
\end{equation}

so

\begin{equation}
\langle \mbox{\boldmath$\tau$}_L \rangle = -\frac{e^2 }{c  R^{5/2} \sqrt{m_e}}  \mbox{\boldmath$\mu$}_{\perp} \times \hat{\mbox{\boldmath$L$}}
\label{torqueporbdtemm7}
\end{equation}

is the average torque on the proton orbit in the electron rest frame due to the proton motion through the electron intrinsic magnetic field.

Using (\ref{Leval}) for the magnitude of \(\mbox{\boldmath$L$}\) this becomes

\begin{equation}
\langle \mbox{\boldmath$\tau$}_L \rangle = \langle\dot{\mbox{\boldmath$L$}}\rangle = \mbox{\boldmath$L$} \times \frac{e}{c R^{3} m_e} \mbox{\boldmath$\mu$}_{\perp} 
\label{torqueporbdtemm12}
\end{equation}

The perpendicular component notation may be dropped since it is defined relative to the orbital angular momentum vector, and replacing the intrinsic magnetic moment with its equivalent in terms of spin results in 

\begin{equation}
\langle\dot{\mbox{\boldmath$L$}}\rangle = -\mbox{\boldmath$L$} \times \frac{e}{c R^{3} m_e} \frac{g e s}{2 m_e c} \mbox{\boldmath$\hat{s}$} = -\mbox{\boldmath$L$} \times \mbox{\boldmath$\omega$}_L
\label{torqueporbdtemm15}
\end{equation}

The precessional angular velocity, averaged over an orbit, of the orbital angular momentum may now be identified to be

\begin{equation}
\mbox{\boldmath$\omega$}_L = \frac{e^2}{c^2  R^{3} } \frac{g s}{2 {m_e}^2} \mbox{\boldmath$\hat{s}$} 
\label{omegaL}
\end{equation}

\subsection{Motion of the Spin Angular Momentum}

In the electron rest frame the torque on the spin is given by 

\begin{equation}
\mbox{\boldmath$\tau$}_s = \mbox{\boldmath$\mu$} \times
\mbox{\boldmath$B$} =  -\frac{g e}{2 m_e c} \mbox{\boldmath$s$} \times
\mbox{\boldmath$B$}
\label{torqueespindtporb}
\end{equation}

where \( \mbox{\boldmath$B$} \) is the magnetic field at the electron, in the electron rest frame, due to the proton orbital motion around the electron, given by

\begin{equation}
\mbox{\boldmath$B$} = \frac{e}{c R^3}\mbox{\boldmath$v$}_p \times \mbox{\boldmath$r$}
\label{Bduetoprot}
\end{equation}

where \( \mbox{\boldmath$v$}_p \) is the proton velocity in the electron rest frame and \( \mbox{\boldmath$r$} \) is the vector from the proton to the electron.

Substituting for \( \mbox{\boldmath$B$} \)  and with \( \mbox{\boldmath$r$}_p = -\mbox{\boldmath$r$} \),  (\ref{torqueespindtporb}) becomes

\begin{equation}
\mbox{\boldmath$\tau$}_s = -\frac{g e}{2 m_e c} \mbox{\boldmath$s$} \times
\left( \frac{e}{c R^3 m_p}\mbox{\boldmath$L$}_p  \right)
\label{muecrossBL2}
\end{equation}

In general  \( \mbox{\boldmath$r$}_p \equiv -\mbox{\boldmath$r$}  \), and also \( \mbox{\boldmath$v$}_p \approx -\mbox{\boldmath$v$}\).  Substituting for \(\mbox{\boldmath$L$}_p\) using (\ref{Lpdef5}) and within the scope of applicability of our approximation we thus obtain

\begin{equation}
\mbox{\boldmath$\tau$}_s  = -\frac{g e^3  s}{2 c^2 {m_e}^{3/2} R^{5/2}} \hat{\mbox{\boldmath$s$}} \times
 \hat{\mbox{\boldmath$L$}} 
\label{muecrossBL4}
\end{equation}

Using this and the expression of (\ref{Thomas_av5}) for \(\mbox{\boldmath$\omega$}_{\text{T}}\),  Equation (\ref{spintransform2}) now becomes

\begin{equation}
\dot{\mbox{\boldmath$s$}} =  -\frac{g e^3  s}{2 c^2 {m_e}^{3/2} R^{5/2}} \hat{\mbox{\boldmath$s$}} \times
 \hat{\mbox{\boldmath$L$}}  - \frac{e^3 s}
{2 c^2 {m_e}^{3/2}R^{5/2}}\hat{\mbox{\boldmath$L$}} \times \hat{\mbox{\boldmath$s$}}
\label{Ns2}
\end{equation}

or

\begin{equation}
\dot{\mbox{\boldmath$s$}} = -\mbox{\boldmath$s$} \times
 \mbox{\boldmath$\omega$}_s = -\mbox{\boldmath$s$} \times \left( \frac{g}{2} - \frac{1}{2}\right) \left( \frac{e^3}
{c^2 {m_e}^{3/2}R^{5/2}} \right) \hat{\mbox{\boldmath$L$}}
\label{Ns5}
\end{equation}

and so the angular velocity of the spin axis around the orbital angular momentum is

\begin{equation}
\mbox{\boldmath$\omega$}_s = \left( \frac{g}{2} - \frac{1}{2}\right) \left( \frac{e^3}
{c^2 {m_e}^{3/2}R^{5/2}} \right) \hat{\mbox{\boldmath$L$}}
\label{omegaslf}
\end{equation}

The leading factor on the right will be recognized as the celebrated Thomas factor that becomes \(\frac{1}{2}\) with \( g=2 \), successfully explaining the factor of \(\frac{1}{2}\) in the anomalous Zeeman effect.

\subsection{Evaluation of the Constancy of the Total Angular Momentum}

It is of interest to determine whether within the present model it is possible to have motions of the spin and orbital angular momentum vectors where the total mechanical angular momentum is a constant of the motion.  It is to be noted that Thomas, using the same model, found that the total secular, \textit{i.e.} orbit averaged, angular momentum is a constant of the motion in the presence of the ``relativity precession'' that now bears his name. The present analysis does not agree with that of Thomas, however.  The difference may be accounted for \cite{LushY9b} as being due to that Thomas did not include in his analysis the ``hidden momentum'' of a magnetic dipole in an electric field.  This could not reasonably be expected of Thomas given that the necessity of accounting for the hidden momentum in the equation of motion of a magnetic dipole was not recognized until the latter half of the twentieth century.

%\begin{figure}
%	\centering
%		\includegraphics[width=0.4\textwidth]{L_s_around_J.eps}
%	\caption{If \(\mbox{\boldmath$L$}\) and \(\mbox{\boldmath$s$}\) precess around %\(\mbox{\boldmath$J$}\), constancy of \(\mbox{\boldmath$J$}\) requires that %\(\mbox{\boldmath$L$}\) and \(\mbox{\boldmath$s$}\) precess with equal angular %velocity. (After \protect\cite{Anderson:ModPhys}).}
%	\label{fig:L_s_around_J}
%\end{figure}

For constant total angular momentum and in accordance with our model that spin and orbital angular momenta magnitudes are constant, it is required that 

\begin{equation}
\frac{d\mbox{\boldmath$J$}}{dt} = \mbox{\boldmath$L$} \times \mbox{\boldmath$\omega$}_L +
\mbox{\boldmath$s$} \times \mbox{\boldmath$\omega$}_{s} = 0
\label{totangeom}
\end{equation}

where \(\mbox{\boldmath$J$} = \mbox{\boldmath$L$} + \mbox{\boldmath$s$}\). This may be rewritten using Eqs. (\ref{omegaL}) and (\ref{omegaslf})  as

\begin{equation}
\mbox{\boldmath$L$} \times \omega_L \hat{\mbox{\boldmath$s$}} +
\mbox{\boldmath$s$} \times \omega_s \hat{\mbox{\boldmath$L$}} = 0
\label{constJcond}
\end{equation}

or

\begin{equation}
(L\omega_L -  s\omega_s)
\hat{\mbox{\boldmath$s$}} \times  \hat{\mbox{\boldmath$L$}} = 0
\label{constJcond4}
\end{equation}

For non-aligned spin and orbital angular momenta, this leads to allowed orbital angular momentum given by

\begin{equation}
L = \frac{\omega_s}{\omega_L} s
\label{constJcond6}
\end{equation}

Substituting for \( \omega_L\) and \(\omega_s\) from (\ref{omegaL}) and (\ref{omegaslf}) and reducing yields 

\begin{equation}
L =  \left( \frac{g}{2} - \frac{1}{2}\right) \left(\frac{2 e {m_e}^{1/2} R^{1/2}}{g}\right) 
\label{conditiononL3}
\end{equation}

Applying the expression of Equation (\ref{Leval}) for \(L\) for circular orbits yields

\begin{equation}
L =  \left( \frac{g}{2} - \frac{1}{2}\right) \left(\frac{2 L}{g}\right) 
\label{conditiononL4}
\end{equation}

which requires for nonzero \( L \) that

\begin{equation}
g =  g - 1 
\label{conditiononL6}
\end{equation}

in order for constancy of the vector total angular momentum to be achieved.  This is a contradiction, for all finite values of \(g\).  Therefore, there exist no radii where angular momentum is constant, for circular orbits, where the spin and orbital angular momenta are not either parallel or antiparallel.  It is also clear from the analysis above that were it not for Thomas precession, total angular momentum constancy would occur within our model for circular orbits for any gyromagnetic factor and spin-orbit relative orientation, and for any orbit radius.   

Muller \cite{Muller:1992} shows that Thomas precession may be considered to arise from a physical torque.  The torque exists absent an externally applied magnetic field.  We see the effect of this torque here, as the impossibility of total mechanical angular momentum constancy for non-parallel spin and orbital angular momenta, for a spinning electron but spinless proton, even absent an externally applied field and associated torque.

It is easy to evaluate specifically what is the total angular momentum deviation from constancy due to Thomas precession.  It was established above that in the presence of Thomas precession \(L \omega_L  -  s \omega_s  \neq 0 \),
regardless of the value of the gyromagnetic factor.   Evaluating this quantity directly obtains

\begin{equation}
L \omega_L  -  s \omega_s  = \left( \frac{1}{2}\right) \left( \frac{e^3 s}
{c^2 m_e^{3/2}R^{5/2}} \right) 
\label{magdJdt11}
\end{equation}

is the rate of change of angular momentum due to Thomas precession.

\section{Motion of the Total Angular Momentum}

Having established that the vector total secular angular momentum cannot be unvarying for nonaligned spin and orbital angular momenta, the nature of the motion of the total secular angular momentum is investigated.

\subsection{Constancy of Total Angular Momentum Magnitude}

It is first considered whether the total angular momentum magnitude may be constant.  The total angular momentum magnitude is constant if \(d(J^2)/dt=0 \), where 

\begin{equation}
J^2 = \mbox{\boldmath$J$} \cdot \mbox{\boldmath$J$} = L^2 + s^2 + 2 \mbox{\boldmath$L$} \cdot \mbox{\boldmath$s$}
\label{Jsqddef}
\end{equation}

The orbital and spin angular momenta magnitudes are constant, so \(d(J^2)/dt=0 \) is equivalent to

\begin{equation}
\dot{\mbox{\boldmath$L$}} \cdot \mbox{\boldmath$s$} = -\mbox{\boldmath$L$} \cdot \dot{\mbox{\boldmath$s$}}
\label{dJsqddteq0}
\end{equation}

or

\begin{equation}
(\mbox{\boldmath$L$} \times \mbox{\boldmath$\omega$}_L) \cdot \mbox{\boldmath$s$} = -\mbox{\boldmath$L$} \cdot (\mbox{\boldmath$s$} \times \mbox{\boldmath$\omega$}_s)
\label{dJsqddteq0_1}
\end{equation}

which, by the properties of the scalar triple product can be rewritten as

\begin{equation}
\mbox{\boldmath$L$} \times \mbox{\boldmath$\omega$}_L \cdot \mbox{\boldmath$s$} = \mbox{\boldmath$L$} \times \mbox{\boldmath$\omega$}_s \cdot \mbox{\boldmath$s$} 
\label{dJsqddteq0_3}
\end{equation}

Taking account that the spin and orbit precess around each other obtains

\begin{equation}
\mbox{\boldmath$L$} \times \omega_L \hat{\mbox{\boldmath$s$}} \cdot \mbox{\boldmath$s$} = \mbox{\boldmath$L$} \times \omega_s \hat{\mbox{\boldmath$L$}} \cdot \mbox{\boldmath$s$} 
\label{dJsqddteq0_3b}
\end{equation}

Both sides of this equation are identically zero by the properties of the scalar triple product, so the requirement of Eq. (\ref{dJsqddteq0}) is satisfied for all \(\mbox{\boldmath$L$} \) and \( \mbox{\boldmath$s$}\).  The magnitude of the total angular momentum is thus a constant of the motion for all relative magnitudes and orientations of \(\mbox{\boldmath$L$} \) and \( \mbox{\boldmath$s$}\), and for all electron-proton separations.

\subsection{Angular Velocity of the Total Angular Momentum}

Next it is evaluated what is the angular velocity of the total angular momentum. Given that \(J\) is constant, the time derivative of the total angular momentum can be written generally as

\begin{equation}
\mbox{\boldmath$L$} \times \omega_L \hat{\mbox{\boldmath$s$}} +
\mbox{\boldmath$s$} \times \omega_s \hat{\mbox{\boldmath$L$}} = \mbox{\boldmath$J$} \times \mbox{\boldmath$\omega$}_J
\label{dJbydt2a}
\end{equation}

where \(\omega_L\) and \(\omega_s\) are constant (as given by Eqs. (\ref{omegaslf}) and  (\ref{omegaL}) ) for a fixed orbit radius, and where \( \mbox{\boldmath$\omega$}_J \) is an angular velocity. This may be written alternatively as   

\begin{equation}
\mbox{\boldmath$J$} \times \omega_L \hat{\mbox{\boldmath$s$}} +
\mbox{\boldmath$J$} \times \omega_s \hat{\mbox{\boldmath$L$}} = \mbox{\boldmath$J$} \times \mbox{\boldmath$\omega$}_J
\label{dJbydt2a}
\end{equation}

from which

\begin{equation}
\mbox{\boldmath$\omega$}_J = \omega_L \hat{\mbox{\boldmath$s$}} +
\omega_s \hat{\mbox{\boldmath$L$}} 
\label{dJbydt2a}
\end{equation}

and

\begin{equation}
\dot{\mbox{\boldmath$\omega$}}_J = \omega_L \dot{\hat{\mbox{\boldmath$s$}}} +
\omega_s \dot{\hat{\mbox{\boldmath$L$}}} = \frac{\omega_L}{s} \dot{\mbox{\boldmath$s$}} +
\frac{\omega_s}{L} \dot{\mbox{\boldmath$L$}} 
\label{dJbydt2a}
\end{equation}

or

\begin{equation}
\dot{\mbox{\boldmath$\omega$}}_J = \frac{\omega_L}{s} (\mbox{\boldmath$s$} \times \omega_s \hat{\mbox{\boldmath$L$}}) +
\frac{\omega_s}{L} (\mbox{\boldmath$L$} \times \omega_L \hat{\mbox{\boldmath$s$}}) 
\label{dJbydt2a}
\end{equation}

or

\begin{equation}
\dot{\mbox{\boldmath$\omega$}}_J = (\omega_L \omega_s - \omega_s\omega_L ) \hat{\mbox{\boldmath$L$}} \times  \hat{\mbox{\boldmath$s$}} \equiv 0
\label{dJbydt2a}
\end{equation}

The angular velocity of the total angular momentum being stationary and constant in magnitude is indicative that the total orbit-averaged angular momentum precesses around a fixed axis.

The magnitude of the angular velocity of the total angular momentum is

\begin{equation}
\omega_J \equiv [\mbox{\boldmath$\omega$}_J \cdot \mbox{\boldmath$\omega$}_J]^{1/2} = [{\omega_L}^2 + {\omega_s}^2 + 2 \omega_L \omega_s (\hat{\mbox{\boldmath$s$}} \cdot \hat{\mbox{\boldmath$L$}})]^{1/2} 
\label{dJbydt2a}
\end{equation}

from which it is apparent that the angular velocity of the total angular momentum vanishes only for antiparallel spin and orbital angular momenta, and for equal mutual precession frequencies.  That the precession frequencies have differing dependences on the orbit radius, according to Eqs. (\ref{omegaslf}) and  (\ref{omegaL}), raises the question of at what orbital radius do they equate.

\subsection{Equation of the Spin and Orbit Precession Frequencies}

Equating \( \omega_s\) and \( \omega_L \) from Eqs. (\ref{omegaslf}) and  (\ref{omegaL}) in accordance with the discussion immediately above yields 

\begin{equation}
\left( \frac{g}{2} - \frac{1}{2}\right) \left( \frac{e^3}
{c^2 {m_e}^{3/2}R^{5/2}} \right)  = \frac{g e^2 s}{2 c^2 {m_e}^{2} R^{3}} 
\label{equateomegas}
\end{equation}

or

\begin{equation}
R^{1/2}  = \left( \frac{g}{2} - \frac{1}{2}\right)^{-1} \frac{g s}{2 e {m_e}^{1/2}} 
\label{equateomegas2}
\end{equation}

which, with \( g = 2 \) and \(s =\hbar/2 \), yields

\begin{equation}
R  =  \frac{\hbar^2}{ e^2 m_e } 
\label{equateomegas3}
\end{equation}

which may be identified as the radius for the ground state orbit in the Bohr model of hydrogen. This value may also be identified as the expectation value of the electron-proton separation in the Schroedinger model of hydrogen.  The other Bohr radii for energy levels above the ground state can be seen to correspond to conditions where the ratio of the spin to orbital angular momentum precession frequencies is an integer value greater than one.  

The classically-expected unity value of \(g\) yields an infinite radius from Eq. (\ref{equateomegas2}).

\section{Conclusion}

A Rutherfordian atomic model with an \textit{ad hoc} incorporation of an electron intrinsic spin and magnetic moment was examined.  A quasiclassical mechanism of quantization was developed based on the expectation that stable states must exhibit total angular momentum constancy as a condition for being nonradiative. Obtaining a value of zero for the angular velocity of the total orbit-averaged angular momentum required equality of the mutual precession frequencies of the spin and orbital angular momenta, as well as that they are antiparallel. It was then shown that in the present model the Bohr radius is the unique radius where the spin and orbit precession frequencies equate, for conventional values of the electron intrinsic spin and magnetic moment magnitudes.  The principal quantum number of the Bohr model was seen to correspond to orbit radii where the spin precession frequency around the orbit normal is an integer multiple of the orbit precession frequency around the electron spin axis.  In the present day where the essentiality of quantum behavior being nonlocal and nondeterministic has been directly challenged \cite{Christian:2007a},\cite{Christian:2007d}, and based on the results described, it may be warranted to reexamine the potential for relativistic classical physics to describe phenomena that were hitherto thought to be purely and fundamentally quantum mechancial in nature. 

\appendix

\section{}

Some standard quantities that were used in the analysis are derived, in the forms needed and in particular for the case of circular orbits. 

\subsection{Circular Orbits and Derivation of the Bohr Radius}

The relationship between orbit radius and velocity for the electron orbiting a heavy proton under Coulomb attraction is derived.

The magnitude of the Coulomb force, \(F\), acting between two
charged particles of equal charge magnitude, \(e\), and separated by
a distance \(R\), is, in Gaussian units,

\begin{equation}
F = \frac{e^2}{R^2}
\label{coulombforce}
\end{equation}

Suppose the electron is in a circular orbit around
the proton, and that the proton is sufficiently heavier than the electron that we may neglect the difference between the proton position and the true center of mass.  Then balancing the centrifugal
force on the electron with the Coulomb attraction from the
proton yields

\begin{equation}
m_e \frac{{v}^2}{R} = \frac{e^2}{R^2}
\label{Feqmaelec}
\end{equation}

where \(m_e\) and \(v\) are the electron mass and velocity, and \(R\) is both the electron-proton
separation and orbit radius in our approximation. The electron velocity as a function of the orbit radius is thus

\begin{equation}
v =  \frac{e}{\sqrt{m_e R}}, \label{elecvelmagnitude4}
\end{equation}

and the orbital angular momentum is

\begin{equation}
\mbox{\boldmath$L$} = \mbox{\boldmath$r$} \times m_e \mbox{\boldmath$v$} = L \hat{\mbox{\boldmath$L$}}= e\sqrt{m_e R} \hat{\mbox{\boldmath$L$}} \label{Leval}
\end{equation}

To derive the ground-state radius \( R_{\text{B}} \) of the Bohr theory of hydrogen, the angular momentum magnitude is set to the reduced Planck's constant \(\hbar\).  Then from (\ref{Leval})

\begin{equation}
R_{\text{B}} = \frac{\hbar^2}{m_e e^2} \approx 5.3 \times 10^{-9}  \text{cm} \label{bohrradiusdef}
\end{equation}

\subsection{Calculation of Thomas Precession Angular Velocity for Circular Orbits}

Aproximating \( \gamma \) as unity in the leading factor in Eq. (\ref{Thomas_freq1}), the Thomas precession angular velocity for circular orbits is

\begin{equation}
\mbox{\boldmath$\omega$}_{\text{T}} = \frac{1}{2}\frac{\mbox{\boldmath$a$} \times
\mbox{\boldmath$v$}} {c^2} =   \frac{1}{2c^2}\left(-\frac{v^2\mbox{\boldmath$r$}}{R^2} \times \frac{m_e \mbox{\boldmath$v$}}{m_e}\right)  
\label{Thomas_av1}
\end{equation}

or

\begin{equation}
\mbox{\boldmath$\omega$}_{\text{T}} =   -\frac{1}{2c^2}\frac{v^2}{m_e R^2}\mbox{\boldmath$L$} 
\label{Thomas_av2}
\end{equation}

and with \(\mbox{\boldmath$L$}  = m_e v R \hat{\mbox{\boldmath$L$}} \) and \(v \) from (\ref{elecvelmagnitude4}), 

\begin{equation}
\mbox{\boldmath$\omega$}_{\text{T}} = - \frac{v^3}
{2 c^2 R}\hat{\mbox{\boldmath$L$}} 
= - \frac{e^3}
{2 c^2 m_e^{3/2}R^{5/2}}\hat{\mbox{\boldmath$L$}} 
\label{Thomas_av5}
\end{equation}

\subsection{Relationship Between Orbital Angular Momentum in the Laboratory and Electron Rest Frames } 

The orbital angular momentum of the proton in the electron rest frame must be evaluated, \( \mbox{\boldmath$L$}_p \), as it is needed to evaluate (\ref{muecrossBL2}).  In the electron rest frame, the proton velocity is the negative of the electron velocity in the laboratory frame, plus an additional component due to the gyration of the electron rest frame.  That is, 

\begin{equation}
\left(\mbox{\boldmath$v$}_p\right)_{\text{elec}} = \left(-\mbox{\boldmath$v$}_e +  R \omega_{\text{T}} \hat{\mbox{\boldmath$v$}}_e\right)_{\text{lab}} = -\mbox{\boldmath$v$}_e \left(1 -  \frac{R \omega_{\text{T}} }{v}\right)
\label{pvelinerf}
\end{equation}

where here the subscript \(e\) for the electron quantities in the laboratory frame is added for clarity.  At the Bohr radius \( R_{\text{B}} \omega_{\text{T}} / v \approx 2.7 \times 10^{-5} \). Approximating the proton velocity magnitude in the electron rest frame by the electron velocity in the laboratory frame thus introduces an error that is smaller by an order of magnitude than the one incurred by neglecting the motion of the proton around the center of mass.  Also  \( \mbox{\boldmath$r$}_p \equiv -\mbox{\boldmath$r$}_e \).  Then   

\begin{equation}
\mbox{\boldmath$L$}_p = \mbox{\boldmath$r$}_p \times
m_p \mbox{\boldmath$v$}_{p} \approx -\mbox{\boldmath$r$}_e \times
-m_p \mbox{\boldmath$v$}_e 
\label{Lpdef}
\end{equation}

where \( m_p \) is the proton mass.  This can be rewritten in terms of the electron angular momentum, \( \mbox{\boldmath$L$} \), in the laboratory frame as

\begin{equation}
\mbox{\boldmath$L$}_p \approx \frac{m_p}{m_e}\mbox{\boldmath$L$}= m_p R v \hat{\mbox{\boldmath$L$}} = m_p e \sqrt{\frac{R}{m_e}} \hat{\mbox{\boldmath$L$}} \label{Lpdef5}
\end{equation}

\bibliographystyle{plain}

%\bibliography{hydrogen}

\end{document}